\documentclass[12pt]{iopart}
\usepackage{graphicx,hyperref}
\usepackage{amssymb,amsthm}
\usepackage{iopams}
\usepackage[margin=1in]{geometry}
\usepackage{tikz-cd}
\usepackage{caption}

\usepackage{xcolor} 

\begin{document}
\title{Metrics on End-Periodic Manifolds as Models for Dark Matter}
\author{Christopher L Duston$^1$}
\ead{dustonc@merrimack.edu}
\address{$^1$ Department of Physics, Merrimack College, N Andover, MA, USA}

\begin{abstract}
  In this paper we will detail an approach to generate metrics and matter models on end-periodic manifolds, which are used extensively in the study of the exotic smooth structures of $\mathbb{R}^4$. We will present three distinct examples, discuss their associated matter models by solving the Einstein equations, and determine their physical viability by examining the energy conditions. We will also compare one of the models directly with existing models of matter distributions in extragalactic systems, to highlight the viability of utilizing exotic smooth structures to understand the existence and distribution of dark matter.
  \end{abstract}

%
%
%

\section{Introduction}\label{s:Intro}
The understanding of our physical Universe as fundamentally a question of geometry is rooted in the key quantitative tool we have to study it, General Relativity (GR). Not only are space and time inexorably linked into a single geometric structure, which dictates how matter moves through it, but the matter content influences that geometric structure directly. Through the development of GR, as physicists have been amassing evidence for it's physical validity, mathematicians have been studying it's formal structure. These formal structures start with the 4D spacetime manifold, and have now branched out into the study of fiber bundles, spinors, strings, and noncommutative algebras, among many others. All of these areas have very well-developed physical motivations as well as mathematical explorations and rigor.

A particular example of an interesting and productive avenue of mathematical exploration has been that of exotic smooth structures. In brief, an exotic smooth structure of a manifold $M$ is a smooth manifold $M_{\theta}$ for which a map between them $f:M\to M_{\theta}$ can only be found to be continuous, not smooth (for a more extensive introduction to this topic, see \cite{Scorpan-2005,AMB-2007}). This relationship between $M$ and $M_{\theta}$ means they share the same topology (large-scale structure), but are inequivalent under diffeomorphisms (small-scale structure), the gauge symmetry of GR. Therefore, metrics on $M_{\theta}$ would be inequivalent solutions to the Einstein equations, despite the two spacetimes having the same topological structure. Specifically, physical predictions requiring the use of derivatives (which characterizes most predictions in physics) would not match on the two manifolds.

Although a seemingly esoteric phenomena, exotic smooth structure has been extensively studied and marveled at in the mathematical community for many years. The first examples were discovered in dimension 7 in 1956 by John Milnor\cite{Milnor-1956}, with work in dimension 4 occurring the 1980s and 1990s\cite{Freedman-1982, Donaldson-1983,Witten-1994}. At this point, there had been little discussion regarding how these structures might impact physical models of our Universe. One of the major complications was the manner of presentation of the available examples - they were very abstract, lacking sufficient details regarding the local geometry that would be needed to define a metric. Early progress was made by Schileich and Witt \cite{Schleich-Witt-1999}, who used the 7-dimensional Wallach spaces $SU(3)/i_{k,l}(\mathbb{S}^1)$ with an embedding $i_{k,l}:\mathbb{S}^1\to SU(3)$ that winds $\mathbb{S}^1$ around a maximal torus in $SU(3)$ according to the integers $k,l$. Through an analysis of the characteristic classes of these spaces it was known that some sets of these spaces were exotic (that is, homeomorphic but not diffeomorphic), and also that they admitted Einstein metrics. As coset spaces, it was possible to find a metric explicitly, and they constructed a semiclassical model in Euclidean quantum gravity to demonstrate that the inclusion of the exotic structures impacted a physical observable (specifically, the expectation value of volume).

Progress in 4-dimensions took somewhat longer. A key approach was initiated by Salvetti \cite{Salvetti-1989}, who used iterated branched covers of $\mathbb{CP}^2$ to calculate the Seiberg-Witten invariants \cite{Witten-1994} and prove there were an infinite number of exotic smooth structures presented in this way. This branched cover $\pi:M\to N$ construction over a manifold $N$ with metric $g$ easily leads to a metric $\pi^* g$ over a particular cover $M$, and this technique was then used to create a model inspired by Schileich and Witt, but this time in 4-dimensions \cite{Duston-2011}. The findings were similar - the inclusion of exotic smoothness structures impacted a physical observable, at least semiclassically.

It's worth mentioning a parallel track of research to access the physical implications of exotic smooth structure, in which knowledge of the metric is foregone in favor of knowledge of the action \cite{Asselmeyer-1997, Asselmeyer-Maluga-2010, Asselmeyer-Maluga-Rose-2012, Asselmeyer-Maluga-Brans-2015}. Briefly, this approach uses techniques from knot theory and surgery to construct exotic smooth structures, and then considers what happens to the curvature under the various actions on the embedded or immersed submanifolds. Using these techniques, a wide variety of interesting phenomena can be developed - for example, exotic smooth structures can be seen to mimic the inclusion of fermions in spacetime, or predict values for the cosmological constant. This is a formally different approach then we will be taking in the present work, and has been included here primarily for context.

So while this brief overview should make it clear that progress is being made, we have not presented any solid predictions of a particular observation which might be explained by exotic smooth structure. The most important possibility for this is undoubtably that of Brans \cite{Brans-1994}, which has come to be known as The Brans Conjecture. It is based on the idea that so-called ``small exotic $\mathbb{R}^4$'' can by embedded in regular $\mathbb{R}^4$, and photons crossing the boundary would be deflected as they crossed. One consequence of this conjecture is that this effect would be observed as gravitational lensing, but the source of the lensing would not be matter, but this exotic manifold structure. Since excess gravitational lensing is attributed to the existence of dark matter \cite{Freese-2009}, the Brans Conjecture represents an alternative explanation for this excess lensing that does not require new or exotic particles or interactions. Since direct experimental searches for these exotic particles have consistently failed to find evidence for them (at the LHC, for example \cite{Aaboud-etal-2019}), exploring manifestations of Brans conjecture is looking more attractive.

In this paper we will be presenting a spacetime model (with a concrete metric and matter content) for such an exotic $\mathbb{R}^4$. It is based on earlier work \cite{Asselmeyer-Maluga-Brans-2012}, which clarified a technique for finding explicit metrics on end-periodic manifolds \cite{Taubes-1987}. In the present work, we will be focusing on finding a valid matter model and comparing the model to those commonly used when studying dark matter, and we will see that the models are comparable at this initial level of analysis. Specifically, for one particular choice of building block we will find matter distribution matching a singular isothermal sphere.

Beyond this introduction, this paper is organized as follows: \S\ref{s:EndPeriodic} will briefly present the theoretical and mathematical background needed to understand the metric generation procedure on end periodic manifolds, and \S\ref{s:Perspective} will give some context for how this approach fits into the larger question, ``what is the dynamical source of exotic smoothness structures?'' We will then present three specific applications of the approach; one starting with a Freeman-Robertson-Lema\'{i}tre-Walker model (\S\ref{s:E-FLRW}), a Kruksal black hole (\S\ref{s:E-Kruskal}), and a embedded conformal surface (\S\ref{s:ConformalSurface}). Possible matter models associated to these metrics will be discussed in \S\ref{s:E-FLRW-Matter} and \S\ref{s:SurfaceMatterModel}, and we will conclude by briefly summarizing our findings in \S\ref{s:Summary}.

\section{Metrics on End-Periodic Manifolds}\label{s:EndPeriodic}
In this section we present the basic background and tools we are using to generate metrics on end-periodic manifolds, the Z-transformation\footnote{The language ``Z-transformation'' comes from discrete signal processing, but seems an appropriate shorthand in this context.}. This was first presented in \cite{Taubes-1987}, where it was used to prove that the number of exotic smoothness structures on $\mathbb{R}^4$ is uncountably infinite. The approach was finally clarified for the physics community in \cite{Asselmeyer-Maluga-Brans-2012,Asselmeyer-Maluga-2016}, and we take this last reference as our primary starting point.

An end-periodic manifold is constructed with a building block 4-manifold $W$ and a map $i$ that identifies the ends of the building block, $Y=W/i$. The end-periodic manifold is the cover
\[\tilde{Y}=...\cup_N W_{-1}\cup_N W_0\cup_N W_1\cup_N...\]
with projection $\pi:\tilde{Y}\to Y$. We also have a map $T:W_i\to W_{i+1}$ which identifies the copies of the building block (this map is also used to define the end-periodic bundles in the original work of \cite{Taubes-1987}). We will pick the metric on the building block $W$ to be $\hat{g}$, which we first have to extend to $Y\times \mathbb{C}^*$ to keep track of the order of the cover. We will do this by making one of the coordinates complex and denote this metric $\hat{g}_z$, $z\in \mathbb{C}^*=\mathbb{S}^1$. Since this metric is periodic, we can define a transformation
\[\hat{g}_z=\sum_{n=-\infty}^{n=\infty}z^{n}(T^ng),\]
where $g$ is the metric on $Y$ and $T^ng$ is the metric on the block $n$. We can invert this expression in a formal way to determine the metric on each block,
\begin{equation}\label{eq:Inverse}
  T^ng=\frac{1}{2\pi i}\oint_{|z|=s}z^{-n}\hat{g}_z\frac{dz}{z}.
  \end{equation}
This expression is independent of the magnitude of the complex coordinate $|z|=s$ thanks to Cauchy's theorem (note that compared to the conventional Z-transformation we have here exchanged $z\leftrightarrow 1/z$ in these two expressions to match what was done originally in \cite{Taubes-1987}).

The next step is to pick an appropriate building block $W$ and metric $\hat{g}$, and decide which part of the metric is going to be periodic. In \cite{Asselmeyer-Maluga-Brans-2012}, the metric choice was Kruskal and the complex coordinate was chosen to be $z=exp(ir)$ with the Kruskal coordinate $v$ tracking the order of the cover. This choice lead to some singular structure in the form of a Heaviside function, suggesting a dimensional reduction from 4 to 2. This dimensional reduction is associated to the a collapse of the 3D hyperbolic spatial section into a 1D tree in the large curvature limit \cite{Asselmeyer-Maluga-2016}, and may have interesting implications for the quantum theory. In \S\ref{s:E-Kruskal} we will take an alternative approach and discuss the differences. 

\section{Perspective Taken in This Paper}\label{s:Perspective}
The Einstein equation tells us the equality between the matter model of the Universe and the geometry (with the cosmological constant being interpreted either way). The traditional way to solve the Einstein equation is to impose symmetries (or Killing vectors) on this system of equations to simultaneously solve for both the matter and geometry together. However, having done this does not usually furnish a complete description of the matter content. For example, a traditional solution for the FLRW provides the scale factor as a function of the density and pressure of an ideal fluid, but does not yet tell us what the matter \textit{content} of the model is. For the case of FLRW, we often have a mixture of radiation, matter, and cosmological constant (as well as more exotic options, see for example \cite{Myrzakul-2016}). Typically observations are required to determine the actual matter content in a particular relativistic model.

Because of the tool we have available to us in the study of end-periodic manifolds, we are going to be taking a slightly different perspective on solving the Einstein equations. The Z-transformation generates new metrics on end-periodic manifolds, but it does not explicitly affect the matter model. Naturally, changing the metric must impact the matter content since that the system still solves the Einstein equation, but the point is that rather then starting with symmetries of spacetime, we are starting with a (hopefully new) metric, and then solving the Einstein equation. By doing this we will be finding a valid matter model (still likely without knowing the specific matter \textit{content}) for a particular choice of geometry. The natural next steps would be to verify that the matter model is physically reasonable, and then checking to see if it can be used for our particular study - to model gravitational lensing without needing dark matter.


A challenging aspect of this field is understanding what the ``dynamical'' source of this exotic smoothness actually is. As discussed in section \S\ref{s:Intro}, they arise because of the existence of a continuous map
\[f:M\to N,\]
but the lack of a smooth one (diffeomorphism). As such, a metric $g$ on $N$ that solves the Einstein equations cannot be pulled back to a metric $f^*g$ on $M$ that is also a solution to the Einstein equations. In other words, solutions to the field equations on $M$ are non-isometric to solutions on $N$. So how should we go about finding the exotic structures $M$ associated to a particular manifold $N$ with known geometric structures? If we knew what the continuous map $f$ was (or say $C^1$ so the pullback was defined), we could use $f^*g$, but not only are such maps typically not part of the construction of exotic smooth structures, that implies that $f^*g$ would at least not be $C^\infty$, raising questions about how it could solve the field equations at all.

Some clarification might come from considering what happens at the level of the action. For a pair of exotic smooth structures $M_1$ and $M_2$, being non-diffeomorphic means the equation
\[\delta S(g)=\delta (S(g)_{EH}+S(g)_{M})=0\]
for the Einstein-Hilbert action $S_{EH}$ and matter action $S_M$, must have different solutions $g_1$, $g_2$. But what kinds of differences in the action would lead to alternative solutions? First, let us restrict our considerations to cases in which the matter content does not change - that is, $S(g_1)_M=S(g_2)_M$. This is consistent with our perspective in this paper, since one could consider the matter action to be independently verified by observations (in the specific case of gravitational lensing, by mass-to-light ratios). If we are therefore just considering what happens to the Einstein-Hilbert action $S_{EH}$, based on the discussion in the introduction it appears we have two essential cases:

\begin{enumerate}
\item  Explicit differences in the action that are parameterized by the construction of the exotic structure. In other words,
  \[S(g_1)_{EH}-S(g_2)_{EH}=F(geometric~parameters).\]
  For a specific example of this, we briefly discuss \cite{Asselmeyer-Maluga-Brans-2015}, in which there is an additional term in the action that takes the form
  \[\int_{U(\Sigma)}\bar{\Phi}D^{U(\Sigma)}\Phi dV.\]
  Here $\Sigma$ is an embedded 3-manifold in a neighborhood $U(\Sigma)\subset M$, which is described via a Weierstrass representation with the spinor $\Phi$. What is important about this characterization is that the geometric parameters of the construction of the exotic smooth structure appear as an extra term in the action. Thus, \textbf{the dynamical source of the exotic structure is the geometric construction of the structure} - surgery, knots, framing, handles, etc.
  \item Differences in the action which cannot be explicitly parameterized as above. In short, these are inequivalent solutions to the Einstein equations which do not have any apparent simple parameterization at the level of the action - one would expect them to have different curvature invariants, for example. Since the field equations are only sensitive to local geometry, the fact that they are topologically identical must come from their abstract presentation. Therefore in this case \textbf{the dynamical source of the exotic structure are local minima in the solution space of the Einstein-Hilbert action}, for a fixed topological background. The primary examples for this second category are found in the semiclassical models, such as \cite{Schleich-Witt-1999, Duston-2011, Duston-2017}.
  \end{enumerate}

We are not suggesting these are the only two possibilities, but they contain all known examples (either the action can be presented differently or not) and provide a useful framework to think about the source of exotic smooth structure. In this paper, we are focusing on the second case, and in the future work we will perform a similar analysis for the first case, to bring models in both categories closer to potential observational verification.

\section{Exotic Friedmann-Lema\'{i}tre-Robertson-Walker Metric}\label{s:E-FLRW}
We will first use the Z-transformation, presented in \S\ref{s:EndPeriodic}, on a building block with the Friedmann-Lema\^itre-Robertson-Walker metric,
\[ds^2=-c^2dt^2+a(t)^2\left(\frac{1}{1-kr^2}dr^2+r^2d\Omega^2\right),\]
Due to the factor $z^{-n}$ in the Z-transformation, we will choose our coordinates to be unitless, our metric components (the scale factor $a(t)$, specifically) to carry units, and subsequently set $c=1$. This also means we are not normalizing $k\in\{-1,0,+1\}$ as is typically done in cosmological studies. We will choose the $(t,r)$ part of the metric to be periodic, following \cite{Asselmeyer-Maluga-Brans-2012}, with the periodic order being tracked by the integer part of the unitless timelike coordinate $n=\left \lfloor t\right \rfloor$ (the lower brackets here indicate the floor function). We also want to complexify the radial coordinate directly, $r\to z\in\mathbb{C}$.


Here our transformation on the radial component of the metric is
\[T^ng_{rr}=\frac{1}{2\pi i}\oint \left(\frac{a(t)^2}{1-kz^2}\right)z^{-n-1}dz,\]
and we need to determine the residue of
\[\frac{1}{z^{n+1}(1-kz^2)}.\]
This expression has singularities at $z=0$ of order $n+1$ and $z^2=1/k$ of order 1. To study the singularity at $z=0$ we can expand in a binomial series and find
\[\frac{1}{z^{n+1}(1-kz^2)}=\sum_{m=0} k^mz^{-(n+1)+2m},\]
which converges for $|kz^2|<1$. 
The term contributing to the residue will be the one with $-(n+1)+2m=-1$, or whenever $m=n/2$ so the value of the residue is $k^{n/2}$.

To get the residue at $z^2=1/k$, we will consider a direct calculation,
\[  \lim_{z\to 1/\sqrt{k}}(z-1/\sqrt{k})\frac{1}{z^{n+1}(1-kz^2)}=-\frac{1}{2}k^{n/2}.\]
So the full result is the combination of these two,
\begin{equation}
  T^ng_{rr}=\frac{1}{2\pi i}\oint \left(\frac{a(t)^2}{1-kz^2}\right)z^{-n-1}dz=a(t)^2(k^{n/2}-\frac{1}{2}k^{n/2})=\frac{1}{2}a(t)^2k^{n/2}.
  \end{equation}

For the time component of the metric, the transformation is
\[T^ng_{tt}=\frac{1}{2\pi i}\oint \left(\frac{-1}{z^{n+1}}dz\right),\]
which is only singular for $n=0$, and in that case has a residue of $-1$. If $n\neq 0$ then the transformation vanishes, and similar to what was found in \cite{Asselmeyer-Maluga-Brans-2012}, we have a reduction in the dimension from $4\to 3$. This could have implications for quantum models using this construction, but for the classical case we will simply assume $t<1$.

The full metric is therefore,

\begin{equation}
  ds^2=\cases{-dt^2+a(t)^2(\frac{1}{2}dr^2+r^2d\Omega^2),& $0\leq t < 1$\\
a(t)^2 (\frac{1}{2}k^{n/2}dr^2+r^2d\Omega^2),& $t>1$\\}
\end{equation}
Notice that although this looks very similar to the original FLRW metric, there is no (obvious) coordinate transformation that will take us back to it for any value of $k$; specifically, this spacetime is no longer maximally symmetric. It is mostly simply related to a monopole metric with a solid angle deficit $\pi(1-k^{-n/2})$. The scalar curvature is
\[R=6\frac{\dot{a}^2}{a^2}+6\frac{\ddot{a}}{a}-\frac{2}{r^2a^2},\]
which looks similar to the $k=0$ FLRW case, but with an extra radial-dependent term. Since the Ricci tensor does not vanish for this spacetime, there is no valid vacuum solution. Even in the case of a static spacetime ($\dot{a}=0$) the curvature does not vanish, due to this additional term.

At first glance, there are some promising signs that this metric might be applicable for gravitational lensing studies. Restricting to the time-invariant case, $\dot{a}\approx 0$, we can consider the Lagrangian for photons:
\[\mathcal{L}=\frac{1}{2}g_{\mu\nu}\dot{x}^{\mu}\dot{x}^\nu\rightarrow 2\mathcal{L}=-\dot{t}^2+\frac{1}{2}a^2\dot{r}^2+a^2r^2(\dot{\theta}^2+\sin^2\theta\dot{\phi}^2).\]
The dots in this expression refer to an independent parameter $\tau$, with the relationship between $\tau$ and $t$ being specified by the equations of motion. Following the usual analysis, we find that the deflection angle is given by
\begin{equation}
  d\phi=\frac{-dr}{r^2\sqrt{\frac{2a^2E^2}{L^2}-\frac{2}{r^2}}}=\frac{-dr}{\sqrt{2}r^2\sqrt{\frac{a^2E^2}{L^2}-\frac{1}{r^2}}},
  \end{equation}
for the constants of motion $L=a^2r^2\dot\phi$ and $E=-\dot{t}$. This goes over to the Schwarzschild answer (divided by $\sqrt{2}$) for an impact parameter of $b=L/(aE)$. 

However, we have not specified a matter model yet; at this stage all know is that it cannot be vacuum. To make progress, we should find a valid matter model for this spacetime by solving the Einstein equations, and then checking the resulting model for it's physical validity. This approach will prove fruitful later; for exotic FRLW we will see this produces matter which violates reasonable energy conditions.

If we start by asking if this metric can be consistent with a perfect fluid matter model, the Einstein equations become (this and most of the following differential geometry calculations were done with SageMath \cite{SageMath}):
\[-\Lambda+3\frac{\dot{a}^2}{a^2}-\frac{1}{r^2a^2}=8\pi \rho,\]
\[\frac{1}{2}\Lambda a^2-\frac{1}{2}\dot{a}^2-a\ddot{a}+\frac{1}{2r^2}=8\pi P_r,\]
\[\Lambda r^2a^2-r^2\dot{a}^2-2r^2a\ddot{a}=8\pi P_{\theta}.\]
From the outset, the chance of finding a consistent solution to these equations that also satisfies a reasonable energy condition seems unlikely. For example, in the $\dot{a}=0$ case, we would need to take something like

\[\rho=-\frac{1}{8\pi}\left(\Lambda+\frac{1}{ r^2a^2}\right)\]
That would imply a spacetime with exotic matter with $\rho<0$, or sufficiently negative cosmological constant for some values in the $(r,a)$-parameter space. In the next section we will explore this solution space extensively and demonstrate that it cannot be found to satisfy any reasonable energy conditions.

\section{A Self-Consistent Matter Model for FLRW}\label{s:E-FLRW-Matter}
It was easy to see in the previous section that simple matter models (such as homogeneous dust, ideal fluid, or scalar field), result in an inconsistent set of field equations. To explore a more complete parameter space, we will consider a more general stress-energy tensor which depends on both $r$ and $t$:

\begin{equation}\label{eq:T_FLRW}
  T^{\mu}_{\nu}=\left( \begin{array}{cccc}
  -T_0(t,r)&&&\\
  &T_1(t,r)&&\\
  &&T_2(t,r)&\\
  &&&T_2(t,r)\end{array}\right)
  \end{equation}

Note that we should not automatically associate the component $T_0$ with the matter density and $T_i$ with fluid pressure - $T_1$ is the radial pressure, and $T_2$ is the angular pressure, which accounts for possible anisotropies. This spacetime is Type I in the Hawking-Ellis classification \cite{Maeda-Martinez-2020}, and see \cite{Pawar-etal-2012} for more on our approach to anisotropic stress-energy tensors.

The only symmetry we are imposing is the $\mathbb{S}^2$ symmetry in the spatial part of the metric. The divergence of the stress-energy tensor is

\begin{eqnarray}
\fl \nabla\cdot T=\left( -\frac{a\left(t\right)^{3} \frac{\partial\,T_0}{\partial t} + {\left(5 \, T_0\left(t, r\right) a\left(t\right)^{2} + 2 \, T_1\left(t, r\right) + 4 \, T_2\left(t, r\right)\right)} \frac{\partial\,a}{\partial t}}{2 \, a\left(t\right)} \right) \mathrm{d} t+\nonumber\\
  + \left( \frac{r \frac{\partial\,T_1}{\partial r} + 2 \, T_1\left(t, r\right) - 2 T_2\left(t, r\right)}{r} \right) \mathrm{d} r\label{eq:DivT}
\end{eqnarray}
By setting $T_1=T_2$, this reduces to the ideal fluid, and the vanishing of this tensor would mean that $T_1$ cannot depend on the radius. This is actually the source of the inconsistency in the Einstein Equations from above; we cannot ask for the three components of the pressure to be isotropic, as in the FLRW case. However, if we keep the spherical symmetry in the pressure and allow the matter model to be anisotropic, we can develop a self-consistent system.

The Einstein equations are then:
\begin{equation}\label{eq:E00}
-\frac{1}{2} \, \kappa T_{0}\left(t, r\right) a\left(t\right)^{2} + \frac{3 \, \frac{\partial}{\partial t}a\left(t\right)^{2}}{a\left(t\right)^{2}} - \frac{1}{r^{2} a\left(t\right)^{2}} = 0
\end{equation}
\begin{equation}\label{eq:E11}
-\kappa T_1\left(t, r\right) - \frac{\frac{\partial}{\partial t}a\left(t\right)^{2}}{a\left(t\right)^{2}} - \frac{2 \, \frac{\partial^{2}}{(\partial t)^{2}}a\left(t\right)}{a\left(t\right)} + \frac{1}{r^{2} a\left(t\right)^{2}} = 0
\end{equation}
\begin{equation}\label{eq:E22}
-\kappa T_2\left(t, r\right) - \frac{\frac{\partial}{\partial t}a\left(t\right)^{2}}{a\left(t\right)^{2}} - \frac{2 \, \frac{\partial^{2}}{(\partial t)^{2}}a\left(t\right)}{a\left(t\right)} = 0
  \end{equation}
with the forth being a duplicate of the third. This last equation suggests that $T_2(r,t)=t_2(t)$, \textit{i.e.} is not a function of $r$. To make further progress, we will propose separable solutions for the rest of the stress-energy tensor,

\[T_0(t,r)=R_0(r)t_0(t), \qquad T_1(t,r)=t_1(t)R_1(r).\]

Starting with the ansatz $t_0(t)=k^2(\dot{a}^2/a^4)$ (note that we started with six variables, $t_0$, $t_1$, $t_2$, $R_0$, $R_1$, and $a$, but only five field equations, so we should expect a degree of indeterminancy, which can be removed by adding a constraint such as this), a solution to the Einstein equations is provided by

\begin{equation}\label{eq:aSol}
  a\left(t\right) = \frac{C_{2} k_{1} + t}{k_{1}},
\end{equation}
\begin{equation}\label{eq:R1Sol}
  R_{0}\left(r\right) = -\frac{2(k_{1}^{2} - 3 \, r^{2})}{k^{2} \kappa r^{2}},
\end{equation}
\begin{equation}\label{eq:R2Sol}
  R_{1}\left(r\right) = \frac{K r^{2} + 2 \, C_{1}}{2 \, r^{2}}
  \end{equation}
\begin{equation}\label{eq:t1Sol}
  t_{0}\left(t\right) = \frac{k^{2}k_1^2}{{\left(C_{2} k_{1} + t\right)}^{4}}
\end{equation}
\begin{equation}\label{eq:t2Sol}
  t_{1}\left(t\right) = -\frac{2}{C_{2}^{2} K k_{1}^{2} \kappa + 2 \, C_{2} K k_{1} \kappa t + K \kappa t^{2}}\end{equation}
\begin{equation}\label{eq:t3Sol}
  t_{2}\left(t\right) = 1/2Kt_1(t)
\end{equation}
where $K$ and $k_1$ are separation constants, $C_1$ and $C_2$ are constants of integration.

By explicit calculation, it can be seen that (\ref{eq:aSol}), (\ref{eq:R1Sol}), (\ref{eq:R2Sol}), (\ref{eq:t1Sol}), (\ref{eq:t2Sol}), and (\ref{eq:t3Sol}) solve the field equations (\ref{eq:E00}), (\ref{eq:E11}), and (\ref{eq:E22}), and ensures the vanishing of (\ref{eq:DivT}), provided we choose the integration constant to be

\[C_1=-\frac{1}{2}Kk_1^2.\]

Of course, the existence of this solution does not mean it has any physical relevance - although we have satisfied energy conservation, there are basic conditions that stress-energy tensors must satisfy for us to consider them viable. In our case we will choose the weak energy condition,

\[T_{\mu\nu}t^{\mu}t^{\nu}\geq 0,\]
for any timelike $t^{\mu}$ as our criteria. Of course, this is the easiest one to satisfy, but without further physical constraints, passing this bar should be considered sufficient for the solution to be considered interesting. For us this expression reads

\[T_0(t^0)^2+\frac{1}{2}a^2T_1(t^1)^2+r^2T_2(t^2)^2+r^2\sin^2\theta T_3(t^3)^2\geq 0.\]

For a timelike vector, this reduces to (this calculation is similar to that which one does for the energy conditions of the ideal fluid)


\[T_0+\frac{1}{2}a^2(T_0+T_1)(t^1)^2+r^2(T_0+T_2)(t^2)^2+r^2\sin^2\theta (T_0+T_3)(t^3)^2\geq 0.\]

So, this condition will surely be satisfied if $T_0$ is positive, and if each of $T_0+T_i$ is also positive. Looking at $T_0=R_0t_0$ (the density), we can see that positivity will occur for $3r^2>k_1^2$. So the density is positive everywhere for large enough values of $r$, a region which we can freely choose by setting $k_1$ to be small enough. The term

\[T_0+T_1=\frac{{\left(C_{2}^{2} - 2\right)} k_{1}^{4} - {\left(C_{2}^{2} - 6\right)} k_{1}^{2} r^{2} + {\left(k_{1}^{2} - r^{2}\right)} t^{2} + 2 \, {\left(C_{2} k_{1}^{3} - C_{2} k_{1} r^{2}\right)} t}{C_{2}^{4} k_{1}^{4} \kappa r^{2} + 4 \, C_{2}^{3} k_{1}^{3} \kappa r^{2} t + 6 \, C_{2}^{2} k_{1}^{2} \kappa r^{2} t^{2} + 4 \, C_{2} k_{1} \kappa r^{2} t^{3} + \kappa r^{2} t^{4}}\]
will be positive everywhere for $\sqrt{2}<C_2<\sqrt{6}$ and $r<k_1$, but that is inconsistent with the positivity condition on the density above. We could perhaps satisfy the null energy conditions (which do not require the density to be independently positive), but looking at the other pieces of the stress-energy tensor,

\begin{equation*}
  \fl T_0+T_2=T_0+T_3=-\frac{{\left(C_{2}^{2} - 6\right)} k_{1}^{2} r^{2} + 2 \, C_{2} k_{1} r^{2} t + 2 \, k_{1}^{4} + r^{2} t^{2}}{C_{2}^{4} k_{1}^{4} \kappa r^{2} + 4 \, C_{2}^{3} k_{1}^{3} \kappa r^{2} t + 6 \, C_{2}^{2} k_{1}^{2} \kappa r^{2} t^{2} + 4 \, C_{2} k_{1} \kappa r^{2} t^{3} + \kappa r^{2} t^{4}},
  \end{equation*}
it is much less clear that even the null conditions can be satisfied. For example, those expressions are negative in the large $t$ regimes, unless $k_1\sim t$ is also large. It would appear that this solution does not satisfy even our relatively relaxed definition of ``physically reasonable'', although it is mathematically self-consistent.

\section{The Exotic Kruskal Metric}\label{s:E-Kruskal}
We will now follow the Z-transformation approach from the previous two sections, but use the Schwarzschild metric in Kruskal-Szekeres coordinates as our building block:

\[ds^2=\left(\frac{32G^3M^3}{r}\right)\exp\left(-\frac{r}{2GM}\right)(-dT^2+dR^2)+r^2d\Omega^2,\]
where the radius is implicitly defined via
\[T^2-R^2=\left(1-\frac{r}{2GM}\right)\exp\left(\frac{r}{2GM}\right).\]
We choose the $(T,R)$ part of the metric to be periodic, tracked by the integer part of the timelike coordinate $n=\left \lfloor T\right \rfloor$. For the complex coordinates, we will simply complexify the radius by setting $r=zr_0$ for $r_0=2GM$, so that $|r/r_0|=s$ is unitless. This should be compared with the original choice in \cite{Asselmeyer-Maluga-Brans-2012}, which was $z=\exp(ir)$. That choice leads to a natural logarithm in the denominator and an essential singularity in the inverse transformation.

\begin{figure}[h]
  \begin{center}
    \includegraphics[scale=0.5]{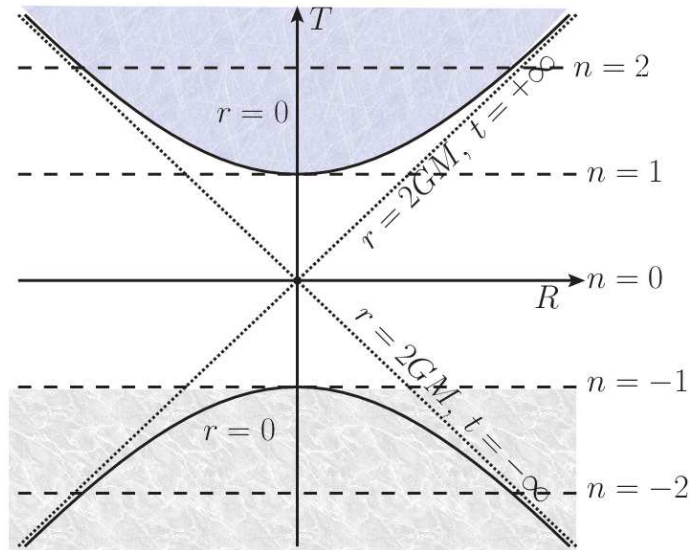}
    \caption{The Exotic Kruskal spacetime diagram with complex radial coordinate. The darkened region $T< -1$ is absent due to a dimensional reduction in these two coordinates - the spacetime is simply a 2-sphere there.}\label{fig:ExoticKruskal}
  \end{center}
\end{figure}

With our choice $r=zr_0$, our inverse transformation (\ref{eq:Inverse}) becomes
\[\frac{16G^2M^2}{2\pi i}\oint \frac{\exp(-z)}{z^{2+n}}dz,\]
and we need to perform a pole analysis on this to determine the value of the integral. Immediately we see that there are no poles for $n< -1$, but for $n\geq-1$ we can determine the pole structure by expanding the exponential:
\[\frac{\exp(-z)}{z^{2+n}}=\sum_{m=0}^{\infty}\frac{\left(-z\right)^m}{m!}\frac{1}{z^{2+n}}=\sum_{m=0}^{\infty}\frac{(-1)^m}{m!}z^{m-n-2}.\]
So for a given value of $n$ we will have a simple pole in the term with $m=1+n$, with residue given by
\[\oint \frac{\exp(-z)}{z^{2+n}}dz=2\pi i\frac{(-1)^{(1+n)}}{(1+n)!},\qquad n\geq -1.\]
As in \cite{Asselmeyer-Maluga-Brans-2012}, in these coordinates we have a dimensional reduction $4\to 2$ for certain values of the timelike coordinate $T<-1$. We can relate this region back to the original spacetime to create a spacetime diagram for this Exotic Kruskal metric, which is shown in figure \ref{fig:ExoticKruskal}. Since the transformation blocks everything $T<-1$, in this model there is no black hole in the past.

The full exotic Kruskal metric is therefore
\begin{equation}\label{eq:ExoticKruskal-TR}
ds^2=\left\{\begin{array}{ll}
\frac{(-1)^{(1+n)}}{(1+n)!}(16G^2M^2)(-dT^2+dR^2)+r^2d\Omega^2&T\geq -1\\
r^2d\Omega^2 & T<-1 \end{array}\right.
\end{equation}

The form of this metric is slightly different from \cite{Asselmeyer-Maluga-Brans-2012}, due to the differences discussed above in the choice of periodic variables. However, it is not amenable to the matter model analysis as we performed in \S\ref{s:E-FLRW-Matter}, because $r$ is defined implicitly in terms of $(T,R)$. More specifically, if we move back to the physical $(t,r)$ coordinates, the metric becomes quite complicated:

\begin{equation}\label{eq:ExoticKruskalMetric}
  \fl g = \frac{re^{\left(\frac{r}{2 G m}\right)}}{Gmb}\left\{ \left(\frac{2Gm}{r}-1\right)\mathrm{d} t^2 - \frac{1}{8}\left(32m^2r^{-1}+12m+r\right)\left( \frac{2 G m}{r} - 1\right)^{-1}\mathrm{d} r^2\right\}  + r^{2} d\Omega^2
  \end{equation}

We will not attempt to solve the field equations for this metric here, but draw some inspiration from (\ref{eq:ExoticKruskal-TR}): our formulation of the Z-transformation produces a conformal transformation on the $\mathbb{R}^2$ part of the spacetime. These types of metrics may have interesting properties in their own right (for example, how might they play a role in conformal gravity \cite{Hooft-2017}?), but we will instead use this as inspiration for another choice of our building block metric in the next section.

\section{A Building Block with an Locally Embedded Conformal Surface}\label{s:ConformalSurface}
Based on the appearance of a locally embedded conformal surface after the Z-transformation in the exotic Kruskal Metric (\ref{eq:ExoticKruskal-TR}), we will consider building blocks of our end periodic manifold $\mathcal{M}$ that produce these same structures more generally. We consider situations when the end period manifold has a metric that looks like

\[ds^2=Z(t,r)(-dt^2+dr^2)+r^2d\Omega^2,\]
where $Z(t,r)$ is the result of some Z-transformation on the $n$th block,

\[Z(t,r)=T^ng=\frac{1}{2\pi i}\oint_{|z|=s}z^{-n}\hat{g}_z\frac{dz}{z}.\]
Mimicking what we've done so far with FLRW and Kruskal, let's start with a building block with topology $\mathbb{R}^2\times \mathbb{S}^2$,

\[ds^2=g(t,r)(-dt^2+dr^2)+r^2d\Omega^2.\]
Further, let us continue to suppose the order of the transformation is tracked by the time coordinate, we will complexify by $r\to z\in \mathbb{C}$, and additionally that our metric function is separable, $g(t,r)=f(t)g(r)$. If we further suppose that $g(r)$ is analytic, it's complexification will have a Laurent series

\[g(z)=\sum_{j=-\infty}^{j=\infty}g_j(z-z_0)^{j}, \qquad g_j=\frac{1}{2\pi i}\oint _{\mathcal{C}}\frac{g(z')dz'}{(z'-z_0)^{j+1}},\]
analytic in a particular region $r_1<|r-r_0|<r_2$ for $r_0=|z_0|$. The transformation will then be

\[Z(t,r)=\frac{f(t)}{2\pi i}\oint_{|z|=r} g(z) z^{-(n+1)}dz=f(t)g_n, \qquad n=\left \lfloor t\right \rfloor.\]
In other words, the effect of the transformation is to grab the $n$th term in the Taylor expansion of the metric on the building block, which will be specified by the time coordinate.

Under this transformation our metric is then

\[ds^2=f(t)g_n(-dt^2+dr^2)+r^2d\Omega^2.\]
If $f(t)$ is sufficiently well-behaved, we can shift the time and radial coordinates;

\[\bar{t}=g_{n} \int_{\lfloor t' \rfloor}^{\bar{t}} f(t')dt',\qquad \bar{t}=0\mbox{ at } \lfloor t' \rfloor,\qquad \bar{t}<1,\]
\[\bar{r}=f(\bar{t})g_n r,\]
\begin{equation}\label{eq:SurfaceMetric}
  ds^2=-d\bar{t}^2+d\bar{r}^2+\left( \frac{\bar{r}}{f(\bar{t})g_n}\right)^2d\Omega^2.
  \end{equation}
This metric is equivalent to the Barriola-Vilenkin monopole \cite{Barriola-Vilenkin-1989},

\[ds^2=-dt^2+d\tilde{r}^2+(1-\Delta)\tilde{r}^2(d\theta^2 + \sin^2\theta d\phi^2),\]
which appears at this level to have an angle deficit that depends on time. When we consider a valid matter model in this case, we will see that the function $f(t)$ must necessarily be constant.

\section{A Self-Consistent Matter Model for $\mathcal{M}$}\label{s:SurfaceMatterModel}
We will start with the same general stress-energy tensor (\ref{eq:T_FLRW}) as we did for FRLW. In this case, the scalar curvature is

\[R=\frac{2 \, g_n f\left(t\right)^{3} - r^{2} \left(\frac{\partial\,f}{\partial t}\right)^{2} + r^{2} f\left(t\right) \frac{\partial^2\,f}{\partial t ^ 2} - 2 \, f\left(t\right)^{2}}{g_n r^{2} f\left(t\right)^{3}}.\]
The curvature vanishes as $r \to \infty$, and is singular at $r=0$. Additionally, for $f(t)=1/g_n$ the curvature is zero, but looking at \ref{eq:SurfaceMetric} we see that this condition simply brings us back to Minkowski space.

The energy conservation equation $\nabla_aT^a_b=0$ is
\begin{equation}\label{eq:Conservation}
  \fl \left( -\frac{2 \, f\left(t\right) \frac{\partial\,T_{0}}{\partial t} + {\left(T_{0}\left(t, r\right) + T_{1}\left(t, r\right)\right)} \frac{\partial\,f}{\partial t}}{2 \, f\left(t\right)} \right) \mathrm{d} t + \left( \frac{r \frac{\partial\,T_{1}}{\partial r} + 2 \, T_{1}\left(t, r\right) - 2 \, T_{2}\left(t, r\right)}{r} \right) \mathrm{d} r=0,
  \end{equation}
and the Einstein equations are

\[-g_n \kappa T_{0}\left(t, r\right) f\left(t\right) + \frac{g_n f\left(t\right)}{r^{2}} - \frac{1}{r^{2}} = 0\]
\[-g_n \kappa T_{1}\left(t, r\right) f\left(t\right) - \frac{g_n f\left(t\right)}{r^{2}} + \frac{1}{r^{2}} = 0\]
\[-\kappa r^{2} T_{2}\left(t, r\right) + \frac{r^{2} \frac{\partial}{\partial t}f\left(t\right)^{2}}{2 \, b f\left(t\right)^{3}} - \frac{r^{2} \frac{\partial^{2}}{(\partial t)^{2}}f\left(t\right)}{2 \, b f\left(t\right)^{2}} = 0\]
It is immediately obvious that $T_0=-T_1$ (or $\rho=P_r$), and further it appears $T_0\sim r^{-2}$ for the first field equation to be satisfied. So, setting

\[T_0(t,r)=\frac{K(t)}{r^2},\]
we can solve for the metric function

\[f(t)=\frac{1}{g_n(1-\kappa K(t))}.\]
However, if we implement these two conditions in the energy conservation equation \ref{eq:Conservation}, we find

\[-\frac{1}{r^2}\frac{\partial K(t)}{\partial t}d \mathrm{t}-\frac{2T_2(t,r)}{r}d \mathrm{r}=0.\]
So $T_2=0$ and $K$ must be a constant, so the density is constant in time. Further, $K\neq 1/\kappa$ or the conformal part of the metric will vanish.  

The radial pressure and density in this model are equal and positive,

\[\rho=P_r=\frac{K}{r^2}>0.\]
So this model satisfies the weak energy condition (although minimally so). The vanishing of the angular pressure $T_2=0$ is a clear sign of the anisotropy, which we will briefly discuss.

For a stress-energy tensor with anisotropic stress we have the generic equation \cite{Pawar-etal-2012}

\[T_{ab}=(\rho+p)u_au_b + \pi_{ab},\]
where the anisotropy can be parameterized as

\[\pi_{ab}=\sqrt{3}S\left(c_ac_b-\frac{1}{3}(u_au_b+g_{ab})\right).\]
Here $S$ is the magnitude of the anisotropy,

\[S=\sqrt{\frac{1}{2}|\pi_{ab}\pi^{ab}|}\]
and $c_a=(0,\sqrt{g_{11}},0,0)$ is a radial vector. In the comoving frame we have

\[T_0^0=\rho,\qquad T_1^1=p+\frac{2}{\sqrt{3}}S,\qquad T_2^2=p-\frac{1}{\sqrt{3}}S,\]
where now $p$ refers to the fluid pressure, rather than the radial pressure $P_r$. The vanishing of $T_2$ simply indicates that the magnitude of the anisotropy is proportional to the fluid pressure,

\[S=\sqrt{3}p.\]
The source of the anisotropy could be self-interactions related to the microscopic details of the matter in question, and demonstrates the issue discussed in \S\ref{s:Perspective} about the lack of specific knowledge of the matter content. We will not attempt to deal with that here, but examples of this type of anisotropy being due to either electromagnetic interactions or scalar fields can be found in \cite{Boonserm-etal-2016}.

This matter distribution is an example of a polytropic equation of state, $P\propto \rho^{\gamma}$, with $\gamma=1$, and is also known as the singular isothermal sphere. By solving the equation of hydrostatic equilibrium, it can be determined that our constant $K$ is related to the central velocity distribution $\sigma_v$ by \cite{Binney-Tremaine-2008}

\[K=\frac{\sigma_v^2}{2\pi G}.\]
This matter model has been traditionally used to describe gravitational lensing in systems such as individual galaxies and X-ray halos \cite{Keeton-2001}. In addition, this mass distribution produces flat rotation curves in spiral galaxies, a classic marker for dark matter \cite{Rubin-etal-1980}.  

We are primarily interested in the viability of this model to describe the excess gravitational lensing usually associated to dark matter. Actually doing that involves reconstructing the mass distribution from a statistical analysis of the deformation and sheer of the 2D projected image of a galaxy cluster on the sky, which would take us quite far afield of our theoretical focus. For a classic introduction to that approach, see \cite{Wambsganss-1998}. It is also true that although the isothermal sphere is often a starting point for the analysis of such astrophysical systems, other models are commonly in use that appear to more accurately describe systems with large amounts of dark matter. Here, we will simply demonstrate that the mass distribution in this model is qualitatively similar to the distributions often found when performing more detailed analysis.

\begin{figure}\begin{center}
    \includegraphics[scale=0.75]{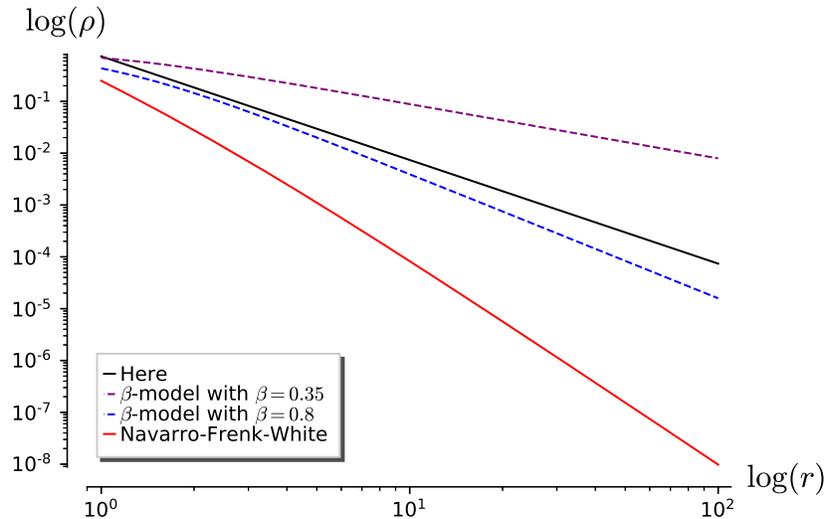}
    \caption{A qualitative comparison between the model in use here (the isothermal sphere) and two alternative models often used to describe the distribution of matter in galactic systems.}\label{fig:ModelComparison}
    \end{center}\end{figure}

To determine the luminous matter in such clusters, X-ray detection of the hot intracluster medium is often used as a proxy. These $\beta$\textit{-models} are parameterized by
\[\rho(r)=\rho_0\left(1+\left(\frac{r}{r_c}\right)\right)^{-3\beta/2},\]
where $\beta\approx 0.65$, but varies for specific cases \cite{Sarazin-1986}. Figure \ref{fig:ModelComparison} shows the minimum and maximum values of $\beta$ in a particular recent survey \cite{Croston-etal-2008}. Dark matter halos in such systems are often described with a Navarro-Frenk-White (MFW) profile, of the form
\[\rho(r)=\rho_0\left\{\frac{r}{r_c}\left(1+\frac{r}{r_c}\right)^{-2}\right\}.\]
We include these comparisons not as a strong argument that we expect it to have similar predictions in terms of lensing as either the $\beta$- or NFW-models, but only as a rough comparison. The fact that our analysis produced one of the most successful models for understanding matter distribution in extragalactic systems should be a strong argument for the validity of the essential approach.

\section{Summary}\label{s:Summary}
In this paper we have explored several examples of metrics on end-periodic manifolds, which are key constructions for the presentation of exotic smooth structure. By necessity, this approach feels ``backwards'' relative to the most common approach to the construction of spacetime models, because the information available about them is quite a bit different. Rather then particular symmetry conditions, or explicit equations of state, all we have to work with is the result of Taubes \cite{Taubes-1987}, that metrics on end-periodic manifolds can be constructed by a periodic transformation on building blocks (the Z-transformation). Once a building block is chosen, the matter model is inferred from the resulting metric, and reality conditions checked.

We have illustrated several examples using this approach, which can be thought of as ranging from the first available, to the first potentially viable. In the case of Exotic-FLRW, while a self-consistent solution to the Einstein equations could be found, the solution did not satisfy reasonable energy conditions. In the case of Exotic Kruskal, the resulting field equations were restrictively complex, although that case served to illustrate that the Z-transformation method (at least in the manner in which we are working with it) produces metrics with embedded conformal surfaces. Using this as a starting point, we generated a simpler building block which resulted in such a metric, and which we used to solve the field equations. The result was a well-known model, the singular isothermal sphere, but with a non-zero stress tensor.

Of course, that matter distribution has been used in astrophysics for years, and is known to fail to describe most observational examples of dark matter. However, the association of what is essentially a monopole metric with a non-vacuum solution to the field equations represents a new connection, forged by considering the underlying spacetime to be an exotic smooth structure, presented as a end-periodic manifold. Perhaps more to the point, it is a slight variation on the original conjecture of Carl Brans - here the exotic smooth structure is not mimicking or generating matter, but rather an exotic smooth structure is shown to contain a standard matter distribution, previously known to the astronomical community. It is our hope that the initial work done here can be developed further to explore the parameter space of exotic smooth structures more completely, perhaps finding models which could stand on their own as explanations for dark matter that do not require the addition of exotic interactions or particles, but only exotic mathematics.

\section*{References}
\bibliographystyle{unsrt}
  \bibliography{myrefs}

\ifx \manfnt \undefined \font\manfnt=logo10 \fi\ifx \METAFONT \undefined \def
  \METAFONT {{\manfnt META}\-{\manfnt FONT}\spac efactor1000 } \fi\ifx \MF
  \undefined \let \MF=\METAFONT \fi\ifx \POSTSCRIPT \undefined \def \POSTSCRIPT
  {{\scshape Post}\-{\scshape Scri pt}\spacefactor1000 } \fi\ifx \MP \undefined
  \def \MP {{\manfnt META}\-{\manfnt POST}\spacefactor1000 } \fi\ifx \noopsort
  \undefined \def \noopsort#1{} \fi\ifx \emdash \undefined \def \emdash{---}
  \fi
\begin{thebibliography}{10}

\bibitem{Scorpan-2005}
Alexandru Scorpan.
\newblock {\em The wild world of 4-manifolds}.
\newblock American Mathematical Society, Providence, RI, 2005.

\bibitem{AMB-2007}
Torsten Asselmeyer-Maluga and Carl~H. Brans.
\newblock {\em Exotic smoothness and physics}.
\newblock World Scientific Publishing Co. Pte. Ltd., Hackensack, NJ, 2007.

\bibitem{Milnor-1956}
John Milnor.
\newblock On manifolds homeomorphic to the {$7$}-sphere.
\newblock {\em Ann. of Math. (2)}, 64:399--405, 1956.

\bibitem{Freedman-1982}
Michael~Hartley Freedman.
\newblock The topology of four-dimensional manifolds.
\newblock {\em J. Differential Geom.}, 17(3):357--453, 1982.

\bibitem{Donaldson-1983}
Simon~K. Donaldson.
\newblock An application of gauge theory to four-dimensional topology.
\newblock {\em J. Differential Geom.}, 18(2):279--315, 1983.

\bibitem{Witten-1994}
Edward Witten.
\newblock Monopoles and four-manifolds.
\newblock {\em Math. Res. Lett.}, 1(6):769--796, 1994.

\bibitem{Schleich-Witt-1999}
Kristin Schleich and Donald Witt.
\newblock Exotic spaces in quantum gravity. {I}. {E}uclidean quantum gravity in
  seven dimensions.
\newblock {\em Classical Quant. Grav.}, 16(7):2447--2469, 1999.

\bibitem{Salvetti-1989}
Mario Salvetti.
\newblock On the number of nonequivalent differentiable structures on
  {$4$}-manifolds.
\newblock {\em Manuscripta Math.}, 63(2):157--171, 1989.

\bibitem{Duston-2011}
Christopher~L. Duston.
\newblock Exotic smoothness in four dimensions and {E}uclidean quantum gravity.
\newblock {\em Int. J. Geom. Methods Mod. Phys.}, 8(3):459--484, 2011.

\bibitem{Asselmeyer-1997}
Torsten Asselmeyer.
\newblock Generation of source terms in general relativity by differential
  structures.
\newblock {\em Classical Quant. Grav.}, 14(3):749--758, 1997.

\bibitem{Asselmeyer-Maluga-2010}
Torsten {Asselmeyer-Maluga}.
\newblock {Exotic smoothness and quantum gravity}.
\newblock {\em Classical and Quantum Gravity}, 27(16):165002, August 2010.

\bibitem{Asselmeyer-Maluga-Rose-2012}
Torsten Asselmeyer-Maluga and Helge Ros\'e.
\newblock On the geometrization of matter by exotic smoothness.
\newblock {\em Gen. Relativity Gravitation}, 44(11):2825--2856, 2012.

\bibitem{Asselmeyer-Maluga-Brans-2015}
Torsten Asselmeyer-Maluga and Carl~H. Brans.
\newblock How to include fermions into general relativity by exotic smoothness.
\newblock {\em Gen. Relativity Gravitation}, 47(3):Art. 47, 27, 2015.

\bibitem{Brans-1994}
Carl~H. Brans.
\newblock Localized exotic smoothness.
\newblock {\em Classical Quant. Grav.}, 11(7):1785--1792, 1994.

\bibitem{Freese-2009}
K.~Freese.
\newblock Review of observational evidence for dark matter in the universe and
  in upcoming searches for dark stars.
\newblock {\em EAS Publications Series}, 36:113–126, 2009.

\bibitem{Aaboud-etal-2019}
M.~Aaboud, G.~Aad, B.~Abbott, D.~C. Abbott, O.~Abdinov, D.~K. Abhayasinghe,
  S.~H. Abidi, O.~S. AbouZeid, N.~L. Abraham, and et~al.
\newblock Constraints on mediator-based dark matter and scalar dark energy
  models using $$ \sqrt{s} $$ = 13 tev pp collision data collected by the atlas
  detector.
\newblock {\em Journal of High Energy Physics}, 2019(5), May 2019.

\bibitem{Asselmeyer-Maluga-Brans-2012}
T.~{Asselmeyer-Maluga} and C.~{Brans}.
\newblock Smoothly exotic black holes.
\newblock In {\em Black Holes: Evolution, Theory and Thermodynamics}, Space
  Science, Exploration and Policies, Physics Research and Technology. NOVA
  Science Publishers, 2012.

\bibitem{Taubes-1987}
Clifford~Henry Taubes.
\newblock Gauge theory on asymptotically periodic {$4$}-manifolds.
\newblock {\em J. Differential Geom.}, 25(3):363--430, 1987.

\bibitem{Asselmeyer-Maluga-2016}
T.~{Asselmeyer-Maluga}.
\newblock Smooth quantum gravity: Exotic smoothness and quantum gravity.
\newblock In {\em At the Frontier of Spacetime}, Fundamental Theories of
  Physics. Springer International Publishing, 2016.

\bibitem{Myrzakul-2016}
Aizhan Myrzakul and Ratbay Myrzakulov.
\newblock On the hojman conservation quantities in frw cosmology, 2016.

\bibitem{Duston-2017}
C.~L. {Duston}.
\newblock {Using cosmic strings to relate local geometry to spatial topology}.
\newblock {\em International Journal of Modern Physics D}, 26:1750033--583,
  2017.

\bibitem{SageMath}
{The Sage Developers}.
\newblock {\em {S}ageMath, the {S}age {M}athematics {S}oftware {S}ystem
  ({V}ersion 9.2)}, 2021.
\newblock {\tt https://www.sagemath.org}.

\bibitem{Maeda-Martinez-2020}
Hideki Maeda and Cristián Martínez.
\newblock {Energy conditions in arbitrary dimensions}.
\newblock {\em Progress of Theoretical and Experimental Physics}, 2020(4), 04
  2020.
\newblock 043E02.

\bibitem{Pawar-etal-2012}
D.D. {Pawar}, V.R. {Patil}, and S.N. {Bayaskar}.
\newblock {Spherically Symmetric Fluid Cosmological Model with Anisotropic
  Stress Tensor in General Relativity}.
\newblock {\em International Scholarly Research Notices}, 2012(965164):10,
  August 2012.

\bibitem{Hooft-2017}
Gerard~'t Hooft.
\newblock Local conformal symmetry in black holes, standard model, and quantum
  gravity.
\newblock {\em International Journal of Modern Physics D}, 26(03):1730006,
  2017.

\bibitem{Barriola-Vilenkin-1989}
Manuel {Barriola} and Alexander {Vilenkin}.
\newblock {Gravitational field of a global monopole}.
\newblock {\em PRL}, 63(4):341--343, July 1989.

\bibitem{Boonserm-etal-2016}
Petarpa Boonserm, Tritos Ngampitipan, and Matt Visser.
\newblock Mimicking static anisotropic fluid spheres in general relativity.
\newblock {\em International Journal of Modern Physics D}, 25(02):1650019, Feb
  2016.

\bibitem{Binney-Tremaine-2008}
James {Binney} and Scott {Tremaine}.
\newblock {\em {Galactic Dynamics: Second Edition}}.
\newblock Princeton Series in Astrophysics, 2008.

\bibitem{Keeton-2001}
Charles~R. {Keeton}.
\newblock {A Catalog of Mass Models for Gravitational Lensing}.
\newblock {\em arXiv e-prints}, pages astro--ph/0102341, February 2001.

\bibitem{Rubin-etal-1980}
V.~C. Rubin, N.~Thonnard, and W.~K. Ford, Jr.
\newblock {Rotational properties of 21 SC galaxies with a large range of
  luminosities and radii, from NGC 4605 /R = 4kpc/ to UGC 2885 /R = 122 kpc/}.
\newblock {\em Astrophys. J.}, 238:471, 1980.

\bibitem{Wambsganss-1998}
Joachim {Wambsganss}.
\newblock {Gravitational Lensing in Astronomy}.
\newblock {\em Living Reviews in Relativity}, 1(1):12, December 1998.

\bibitem{Sarazin-1986}
Craig~L. Sarazin.
\newblock X-ray emission from clusters of galaxies.
\newblock {\em Rev. Mod. Phys.}, 58:1--115, Jan 1986.

\bibitem{Croston-etal-2008}
J.~H. Croston, G.~W. Pratt, H.~Böhringer, M.~Arnaud, E.~Pointecouteau, T.~J.
  Ponman, A.~J.~R. Saderson, R.~F. Temple, R.~G. Bower, and M.~Donahue.
\newblock Galaxy-cluster gas-density distributions of the representative
  xmm-newton cluster structure survey (rexcess).
\newblock {\em Astronomy \& Astrophysics}, 487(2):431–443, Apr 2008.

\end{thebibliography}

\end{document}